\documentclass[12pt]{article}
\usepackage{latexsym,amsmath,amssymb,amsbsy,graphicx}
\usepackage[utf8]{inputenc}
\usepackage[T2A]{fontenc}
\usepackage[space]{cite} 

\makeatletter\def\@biblabel#1{\hfill#1.}\makeatother

\allowdisplaybreaks
\begin {document}

\noindent\begin{minipage}{\textwidth}
\begin{center}

{\Large{On the Influence of “red leak”{} of Light Filters on the Brightness Estimates of Stars of Late Spectral Types Illustrated by the Observations of Rapid Variability of Symbiotic Stars}}\\[9pt]

{\large G.\,E. Nikishev$^{1}$, N.\,A. Maslennikova$^{1a}$, A.\,M. Tatarnikov$^{1}$, K.Yu.\, Parusov$^{1}$, A.A.\, Belinski$^{1}$}\\[6pt]

\parbox{.96\textwidth}{\centering\small\it
$^1$ M.V.Lomonosov Moscow State University, Moscow 119191, Russia. \\
\ E-mail: $^a$maslennikova.na16@physics.msu.ru}\\[1cc] 

\parbox{.96\textwidth}{\centering\small Received: 22.09.2023/ Accepted: 17.10.2023.} 
\end{center}

{\parindent5mm The results of modeling the dependence of the “red leak” {} of photometric filters on various factors (color index $V-R$, luminosity class, interstellar reddening, air mass and PWV) during observations of stars are presented. The error arising from not taking into account the “red leak”{} in the case of filters used on the 0.6-m telescope of the CMO SAI can amount to $0.6^m-0.8^m$ for late stars. Algorithms for reducing observational data are presented for filters $U$ and $B$. The results of observations of the rapid variability of two symbiotic stars CH~Cyg and SU~Lyn with cold components of very late spectral types are presented. For CH~Cyg, rapid variability was detected on both observation dates. Taking into account the “red leak”{} effect, the amplitude in the $B$ band was 0.10$^m$ on November 6, 2019 and 0.19$^m$ on December 15, 2022, with a characteristic variability time of about 20 minutes. For SU~Lyn, no rapid brightness variability was detected in the $B$ band on February 2, 2023 (with an accuracy of 0.003$^m$).

\vspace{2pt}\par}

\vspace{1pt}\par

\vspace{1pt}\par
\end{minipage}

\section*{Introduction}
\mbox{}\vspace{-\baselineskip}

Symbiotic stars are binary systems, in the spectra of which the spectrum of cool stars~--- a red giant - and emission lines with a high ionization potential, typical, for example, for the spectra of planetary nebulae, are simultaneously visible \cite{Boyarchuk}. Thus, one component of the binary system is a red giant with a temperature of 3500~K (sometimes a red supergiant or yellow giant), and the second is a hot subdwarf (a star similar to the central stars of planetary nebulae) with a temperature $T_{hot}>30000$~К (in some systems the hot component is a white dwarf or neutron star). The maximum radiation of the hot component occurs in the far UV band, so it ionizes the substance lost by the cold component. As a result, an extended ionized shell is formed, in which emission lines such as H~I, He~I, He~II, [O~III], etc. are formed. An accretion disk may exist around the hot component of a symbiotic star (for example, \cite{Bath_1982}, \cite{Kafatos_1982}, \cite{Viotti_1982}, \cite{Bollimpalli_2018}). This combination of objects of different types in one system leads to the fact that symbiotic stars exhibit manifestations of activity of different scales (from thousandths of a magnitude to several magnitudes) and at different time periods~--- from several minutes to tens and hundreds of years.

One of the types of variability of symbiotic stars is rapid brightness variability, similar in its parameters to the flicker effect of cataclysmic binaries \cite{Bruch2000}~--- variability with a characteristic amplitude of up to tenths of a magnitude at times from several minutes to an hour. At the same time, the variability can be either periodic (for example, in BF~Cyg \cite{Formiggini2009} or Z~And \cite{Sokoloski1999}), or irregular~--- CH~Cyg \cite{Stoyanov2018}, T~CrB \cite{Dobrotka2010}, CSS~1102 \cite{Maslennikova2022} etc. The origin of variability in the first case is a rotation of the hot component, and in the second, apparently, transient phenomena during the accretion.

The search for and study of the rapid variability of symbiotic stars is currently being carried out photometrically with CCD cameras. Observations are carried out in the spectral range in which the maximum contribution of the variable source is expected. The cold component of the system has a maximum emission in the red or near-infrared region of the spectrum, its emission at characteristic monitoring times (hundreds of minutes) can be considered constant and its contribution drops rapidly with decreasing wavelength. Therefore, observations are usually made in the $U$ or $B$ bands (center wavelength 360~nm or 440~nm, respectively). However, modern CCD detectors have a very wide sensitivity bandwidth, extending from 300-350~nm to 1100-1200~nm. If the used shortwave filter has transmission not only in the working area, but also in the red part of the spectrum~--- the so-called “red leak”{}, then during observations not only photons in the band, for example, $B$, but also several from the long-wave part will be recorded. When observing very red objects (e.g., late M-stars), the effect can be very large even for small (fractions of a percent) amounts of red leak. The effects of red leak were affected, for example, by HST\footnote{https://www.stsci.edu/instruments/wfpc2/Wfpc2\_hand/HTML/W2\_30.html} UV filters and the $u$ filter of the SDSS survey\footnote{https://www.sdss4.org/dr17/imaging/caveats/}. The necessity of taking into account the red leak in symbiotic star observations \cite{Munari2012} is also known. To some extent, the influence of this effect can be reduced by selecting comparison stars of similar spectral types. However, this is not always possible - firstly, such stars may have their own variability, and secondly, stars with the required color indices may simply not be in the field of view of the photometer.

In this paper, we study the impact of the “red leak”{} of the CCD photometer filters of the RC600 telescope of the Caucasus Mountain Observatory of SAI MSU (CMO) on the results of red stars photometry and determine a method for taking this effect into account using the example of processing rapid variability observations of the symbiotic star CH~Cyg and variable star SU~Lyn, suspected of belonging to the class of symbiotic stars, the cool components of which have late spectral types.

\section{Observations}
\mbox{}\vspace{-\baselineskip}

\begin{figure}[h]
\center{
\includegraphics[width=0.5\linewidth]{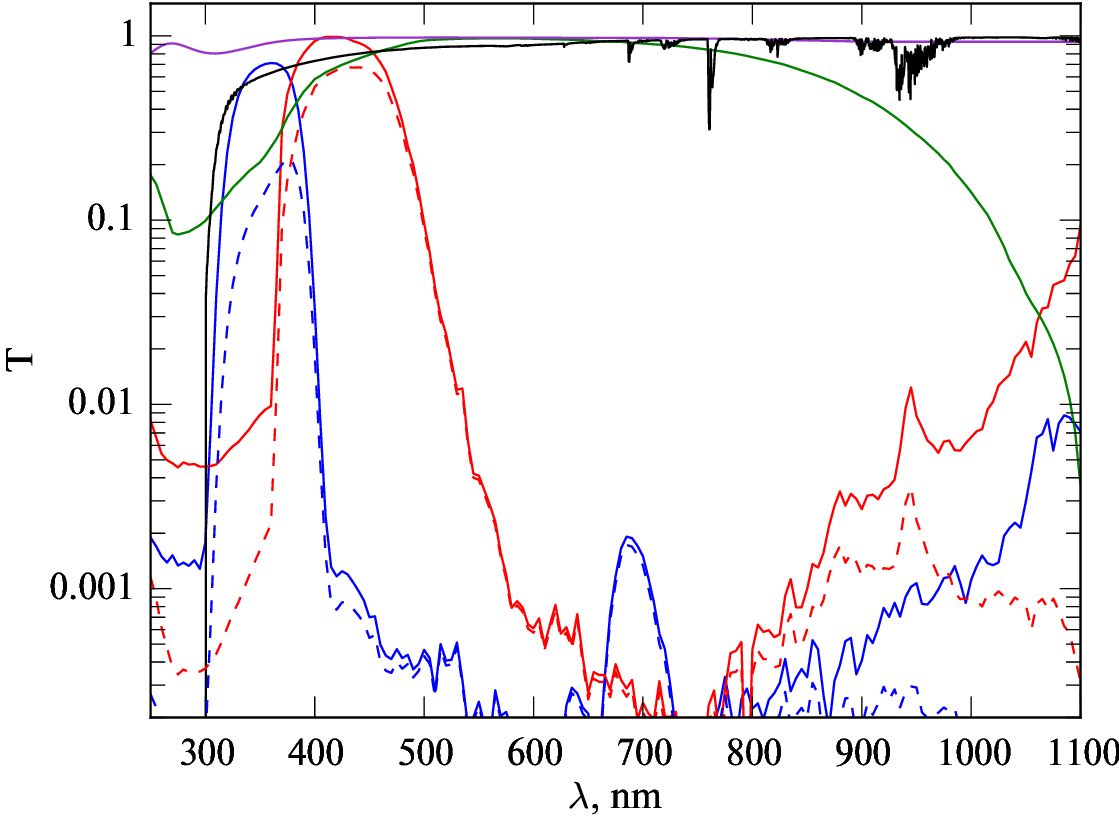}
}
\caption{Transmission curves of the $U$ (blue line) and $B$ (red line) filters, atmosphere (black line), camera entrance window (purple line) and bandwidth of the Andor iKON-L CCD camera (green line). Dashed lines show the response curves of the corresponding instrumental bands (product of filter bandwidth, entrance window and camera sensitivity) 
}
\label{fig:fig01}
\end{figure}

At the SAI CMO \cite{Shatsky2020}, the 60-cm RC600 telescope manufactured by ASA (Austria) has been actively operating since 2019. It is used for a large volume of photometric observations. For this purpose, a CCD photometer \cite{Berdnikov2020} based on the Andor iKON-L CCD matrix (image scale 0.67~arcsec/pixel) is used. The photometer is equipped with a filter wheel containing $UBVR_cI_c$ filters of the Johnson-Cousins photometric system. The transmission curves of all filters are available on the CMO website\footnote{https://arca.sai.msu.ru/filters}. The $R_c$ and $I_c$ bands used in this work will be referred to as $R$ and $I$ in the text. The sensitivity curves of the CCD matrix detector and the transmission curves of the short-wave $U$ and $B$ filters, used since 2019, registered on the Avantes avaspec 2048 spectrophotometer, are shown in Fig.~\ref{fig:fig01}. Also installed on the RC600 are the $g'$, $r'$, $i'$, and $z'$ filters of the SDSS photometric system. The transmission curve of the short-wave $g'$ filter outside the main band is very similar to the $B$ curve, so our results of the ”red leak” research in the $B$ band are also applicable to observations in this filter.

The photometric observations discussed below were carried out at RC600. The symbiotic star CH~Cyg was observed on 06.11.2019 and 15.12.2022 in the $B$ band. The variable SU~Lyn suspected of belonging to symbiotic stars \cite{Ilkiewicz_2022} was observed on 25.02.2023 in the $U$ band. The photometric observations were processed in the standard way (accounting for the bias, dark and flat field, and aperture photometry). The errors were estimated for field stars whose brightness is close to the magnitudes of the objects. For CH Cyg, the single measurement errors on 06.11.2019 were $0.006^m$, on 15.12.2022~--- $0.005^m$, and for SU~Lyn~--- $0.003^m$.

\section{Synthetic photometry}
\mbox{}\vspace{-\baselineskip}

As can be seen from Fig.~\ref{fig:fig01}, the $U$ and $B$ filters currently used on the RC600 telescope have significant red leak. For the $U$ filter it is represented as a narrow peak at 700~nm with a maximum transmittance of about $0.2\%$, while for the $B$ filter it consists of two regions~--- red (in the wavelength range of $550-700$~nm with an average transmittance of $<0.1\%$) and extensive infrared with an average transmittance of about $1\%$.

The final value of red leak is also influenced by the transmission of the Earth's atmosphere and the entrance window of the camera, and the sensitivity of the detector. Fig.~\ref{fig:fig01} shows the Earth's atmospheric transmittance curve for a high-altitude observatory (CMO altitude is about 2100~m above sea level). It was derived from the ESO SkyCalc\footnote{https://www.eso.org/observing/etc/bin/gen/form?INS.MODE=swspectr+INS.NAME=SKYCALC} online calculator for an altitude of 2400~m and water content PWV=2.5~mm, and then adjusted to the altitude and pressure typical of CMO according to \cite{Noll2012}. It can be seen that the influence of the atmosphere on the leakage value can be practically neglected. The sensitivity curve of the detector has slopes in both the UV and IR parts of the spectrum, and affects the considered filters differently. The UV slope of the curve reduces the efficiency of the $U$ filter in its main bandwidth (thereby increasing the relative contribution of red leak). The IR slump of the curve significantly affects the value of the IR part of the $B$ filter leak (see Fig.~\ref{fig:fig01}).

To determine the magnitude of the red leakage contribution to the signal in the short-wavelength instrumental bands, we performed synthetic photometry on the spectra of stars of different spectral types (from O5 to M10) presented in the \cite{Pickles1998} spectra library. The synthetic signal for a star of each spectral type was calculated by integrating the convolution of the spectrum $I(\lambda)$ with the atmospheric transmission curve $A(\lambda)$ and the instrumental band response curves $F(\lambda)$ shown in Fig.~\ref{fig:fig01} (the multiplier $\lambda$ under the integral is used to convert the energy into the number of photons):

\begin{equation}
N_{ph} \sim \int^{\lambda_2}_{\lambda_1} \lambda I(\lambda) A(\lambda) F(\lambda) d \lambda
 \label{eq:flux}
\end{equation}

Then, the value of the red leak (in stellar magnitudes) can be calculated by the formula:

\begin{equation}
\delta {\rm mag} = 2.5 \lg \frac{\int^{1100}_{300} \lambda I(\lambda) A(\lambda) F(\lambda) d \lambda}{\int^{\lambda_2}_{\lambda_1} \lambda I(\lambda) A(\lambda) F(\lambda) d \lambda}
 \label{eq:mag}
\end{equation}

In the numerator, integration is carried out over the entire spectral band, and in the denominator~-- between wavelengths that limit the working band of the filter (for the $U$ filter $\lambda_1=315$~nm, $\lambda_2=470$~nm, for the $B$ filter $\lambda_1=365$~nm, $\lambda_2=575$~nm).

The color indices $V-R$ for each spectrum were also calculated by synthetic photometry. The spectrum of the star A0V was used as a standard, and the filters $V$ and $R$ were considered free of red leakage.

\begin{figure}[h]
\center{
\includegraphics[width=0.5\linewidth]{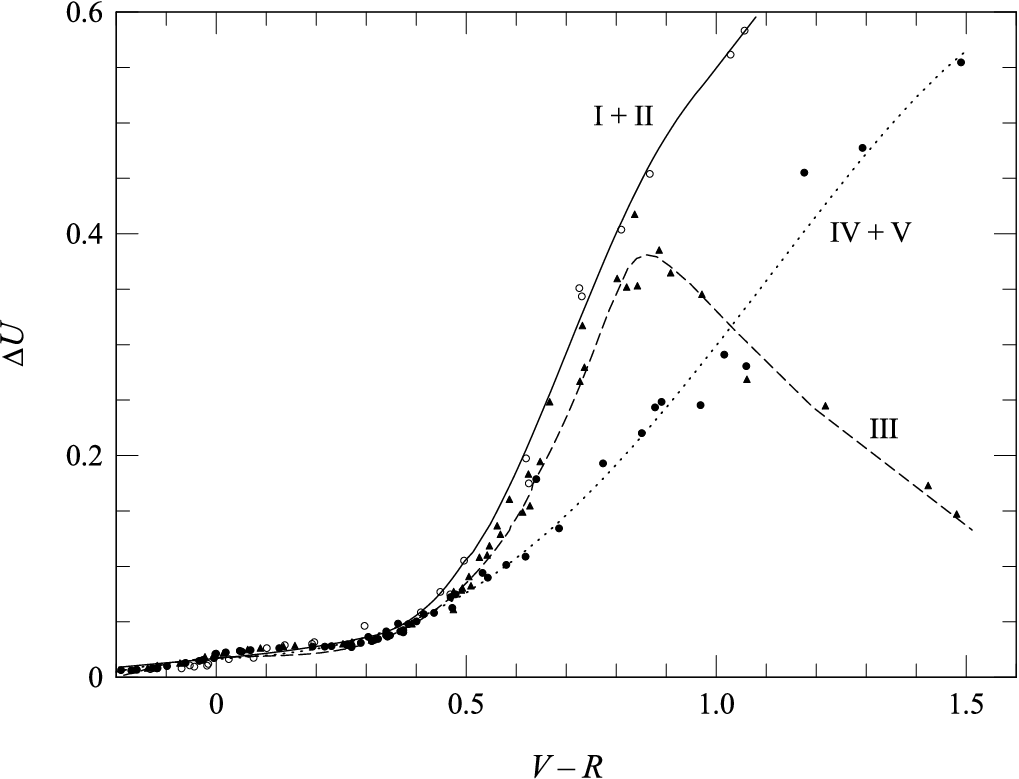}
}
\caption{Dependence of the red leak in the $U$ band on color for stars of different luminosity classes (circles~--- I and II classes, triangles~--- III class, black dots~--- IV and V classes; the corresponding lines show smoothed dependences)
}
\label{fig:fig_dU}
\end{figure}

Fig.~\ref{fig:fig_dU} shows the results of calculations of the red leak value for the $U$ filter. It can be seen that for stars of late spectral types (later than about G0), the single dependence of the leakage value on the color index splits into two~--- for I~-- III classes and IV~-- V classes. In the region $(V-R)\approx0.8$ (K stars), the curves split once more~--- for red giants, the leakage value decreases with increasing color index. Thus, to correctly account for the read leak in the $U$ band, it is necessary to know not only the color of the star (or spectral types), but also the luminosity class.


\begin{figure}[h]
\center{
\includegraphics[width=0.5\linewidth]{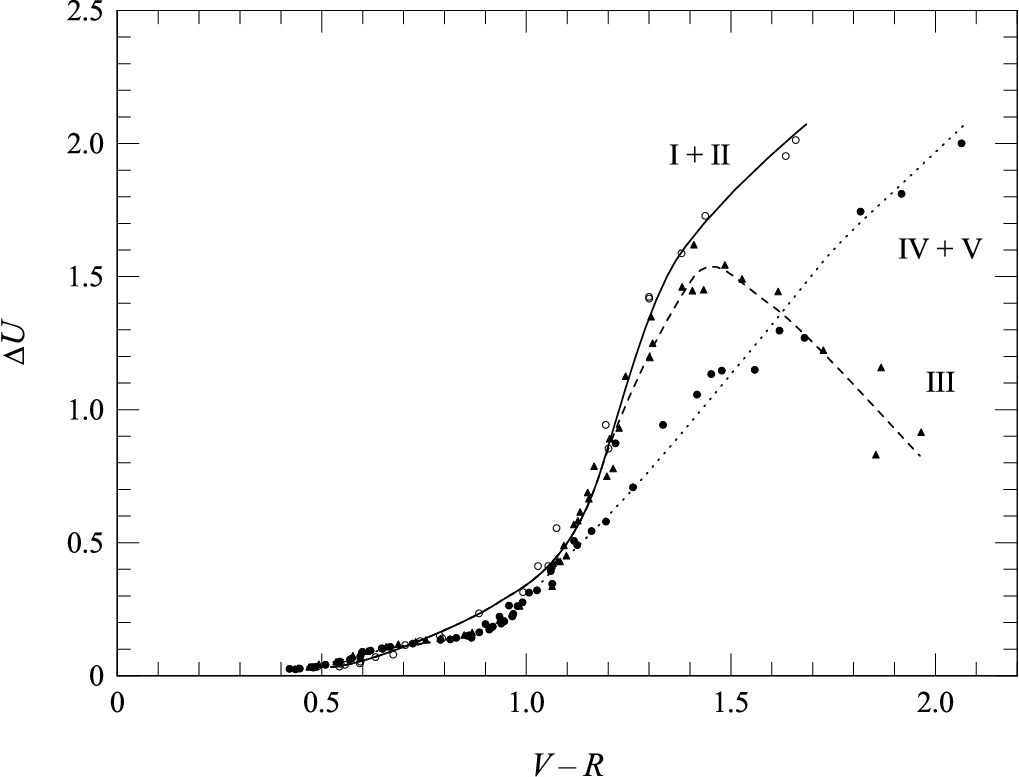}
}
\caption{Dependence of the red leak value in the $U$ band on color for the case of interstellar reddening with color excess $E(B-V)=1$ for stars of different luminosity classes
}
\label{fig:fig_dU_EBV}
\end{figure}

As the calculations show, the magnitude of the red leak is affected not only by the color index but also by the value of the interstellar reddening. Fig.~\ref{fig:fig_dU_EBV} shows the leakage in the $U$ band for the case of $E(B-V)=1$. It can be seen that it is still possible to identify separate color dependences of the leakage for late-type stars of different luminosity classes. The leakage grows rapidly with increasing $V-R$ and reaches $2^m$ (i.e., the leakage contribution to the total signal is 6 times larger than the useful signal).

\begin{figure}[h]
\center{
\includegraphics[width=0.5\linewidth]{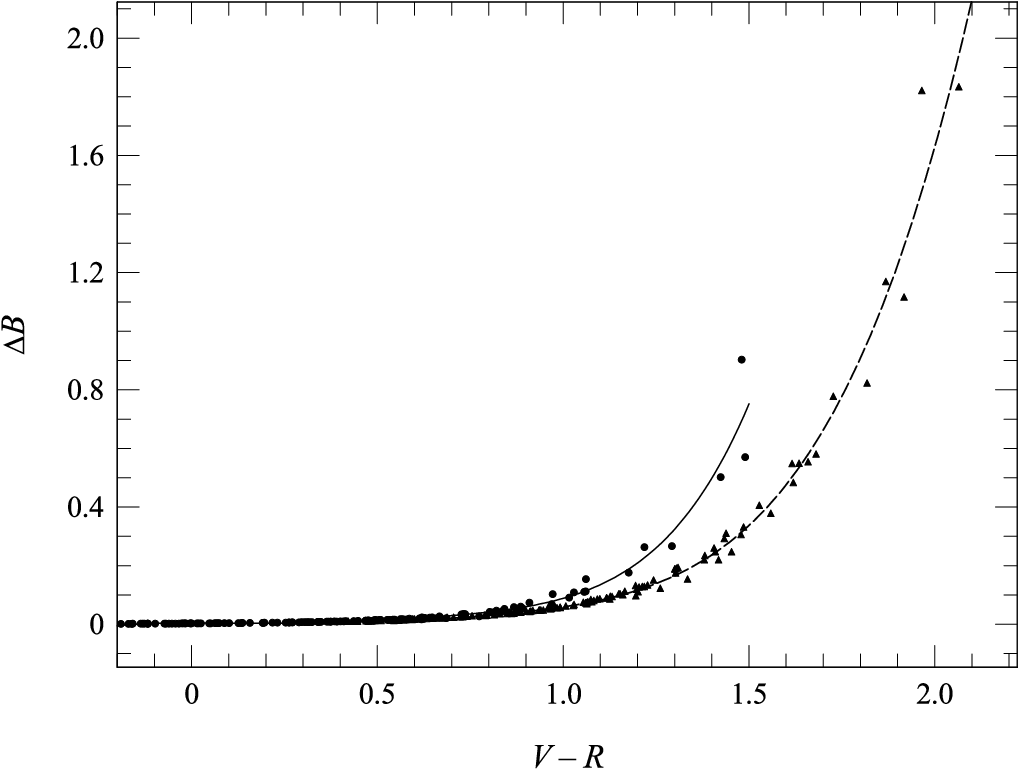}
}
\caption{Dependence of the red leak value in the $B$ band on color for stars of luminosity classes I~-- V in the case of no interstellar reddening (dots and solid line) and in the case of excess color $E(B-V)=1$ under the normal reddening law (triangles and dashed line)
}
\label{fig:fig_dB}
\end{figure}

In the $B$ band (Fig.~\ref{fig:fig_dB}), there is practically no separation of the dependence by luminosity classes~--- the leakage value slowly increases from 0 for hot stars to $0.05^m$ for stars of spectral type $K0-K1$, after which it begins its sharp increase, and at $V-R\sim 1.5$ it can exceed $0.5^m$. Interstellar reddening, just as in the case of the $U$ band, significantly increases the leakage value for stars of the same spectral types.

\begin{figure}[h]
\center{
\includegraphics[width=0.5\linewidth]{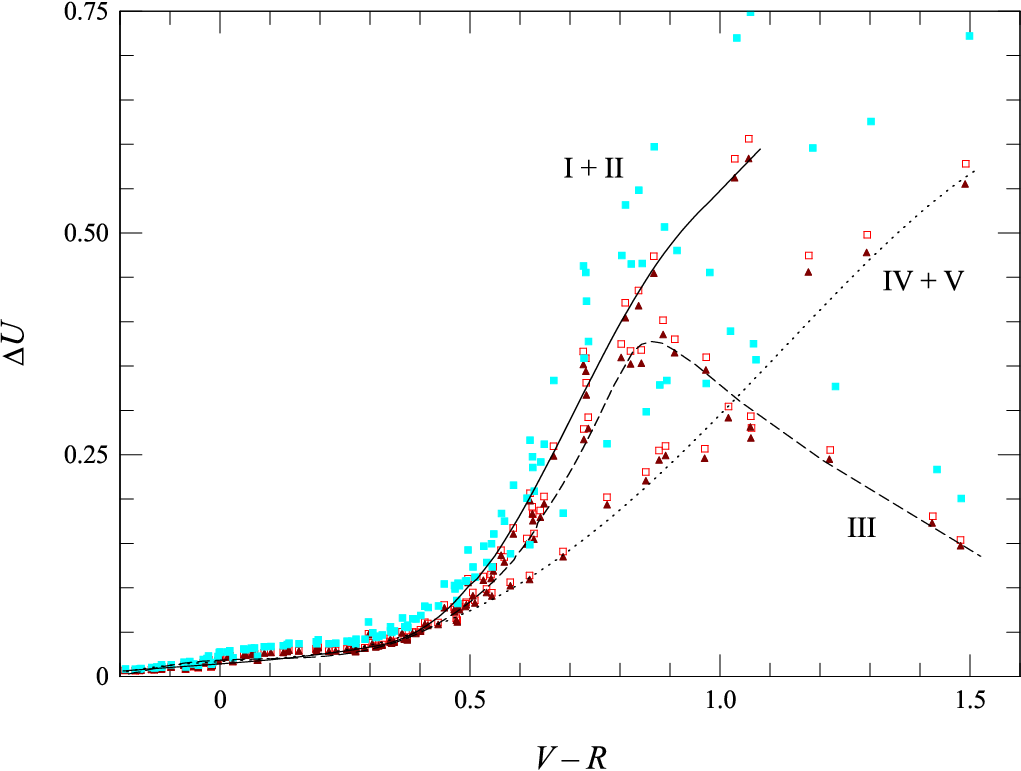}
}
\caption{Dependence of the red leak value in the $U$ band on the color of stars for different values of airmass at $PWV=2.5$~mm: $M_z=1$~--- black triangles, $M_z=1.15$~--- red squares, $M_z=2$~--- blue squares. The lines show the positions of the sequences highlighted in Fig.~\ref{fig:fig_dU}
}
\label{fig:fig_dU_alt}
\end{figure}

\begin{figure}[h]
\center{
\includegraphics[width=0.5\linewidth]{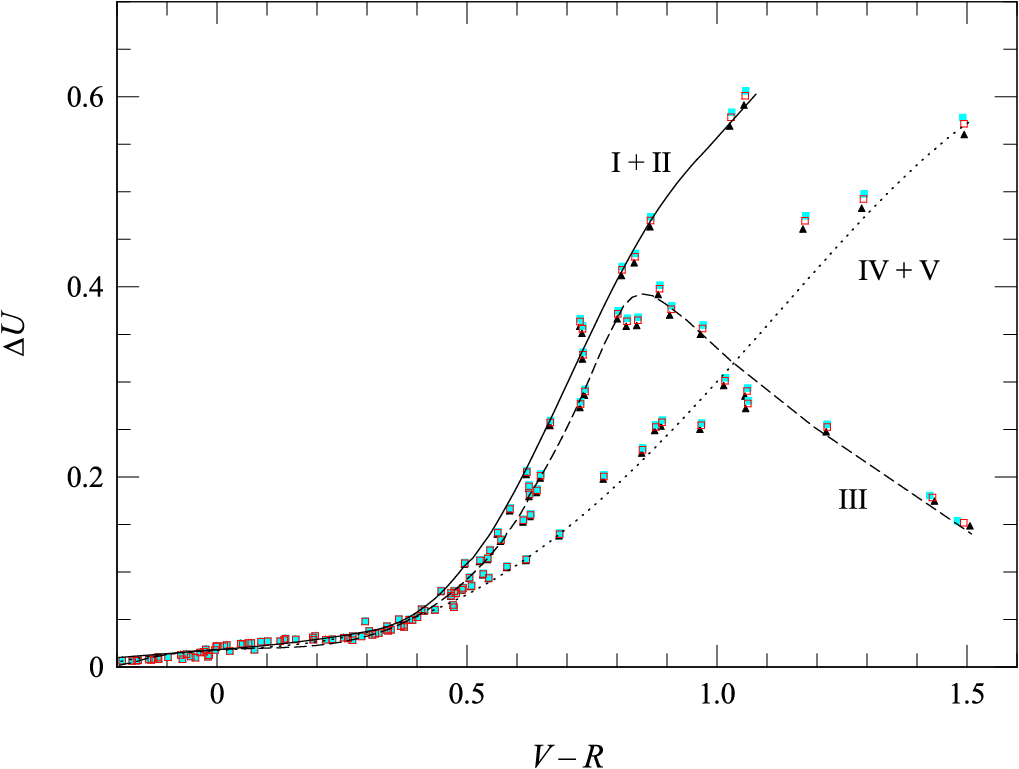}
}
\caption{Dependence of the red leak value in the $U$ band on the color of the stars for different values of $PWV$ at $Mz=1.15$: $PWV=2.5$~mm~--- black triangles, $PWV=10$~mm~--- red squares, $PWV=30$~mm~--- blue squares. The lines show the positions of the sequences highlighted in Fig.~\ref{fig:fig_dU}
}
\label{fig:fig_dU_pwv}
\end{figure}

\begin{figure}[h]
\center{
\includegraphics[width=0.5\linewidth]{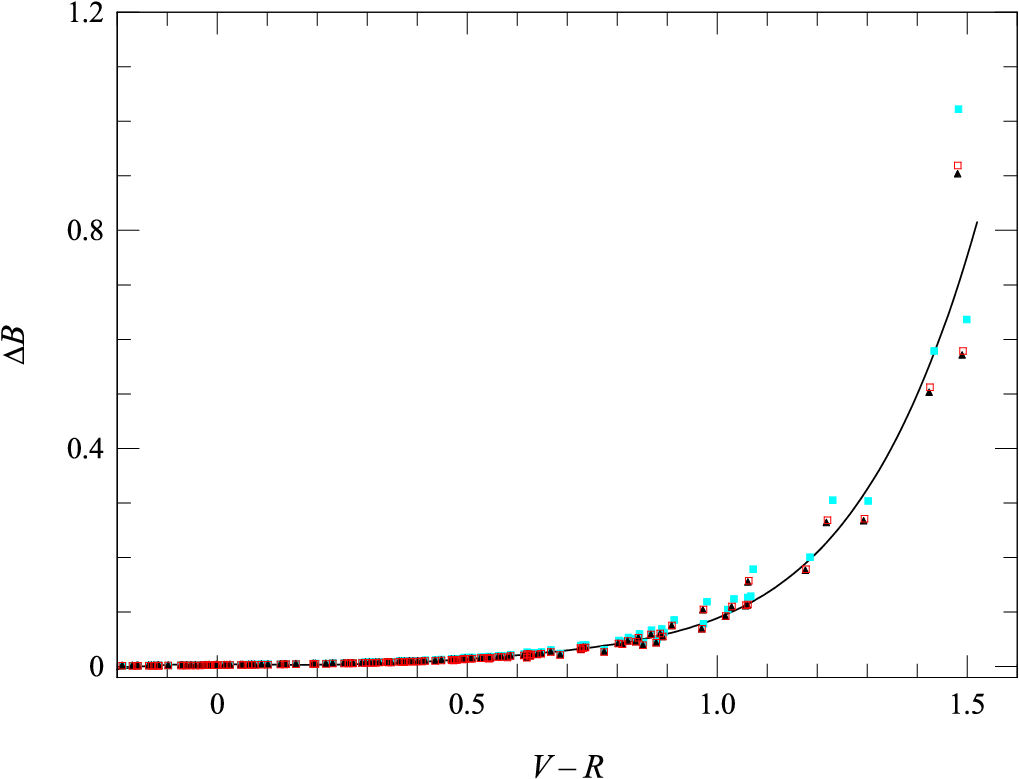}
}
\caption{Dependence of the red leak value in the $B$ band on the color of stars for different values of airmass at $PWV=2.5$~mm. The designations are the same as in Fig.\ref{fig:fig_dU_alt}
}
\label{fig:fig_dB_alt}
\end{figure}

\begin{figure}[h]
\center{
\includegraphics[width=0.5\linewidth]{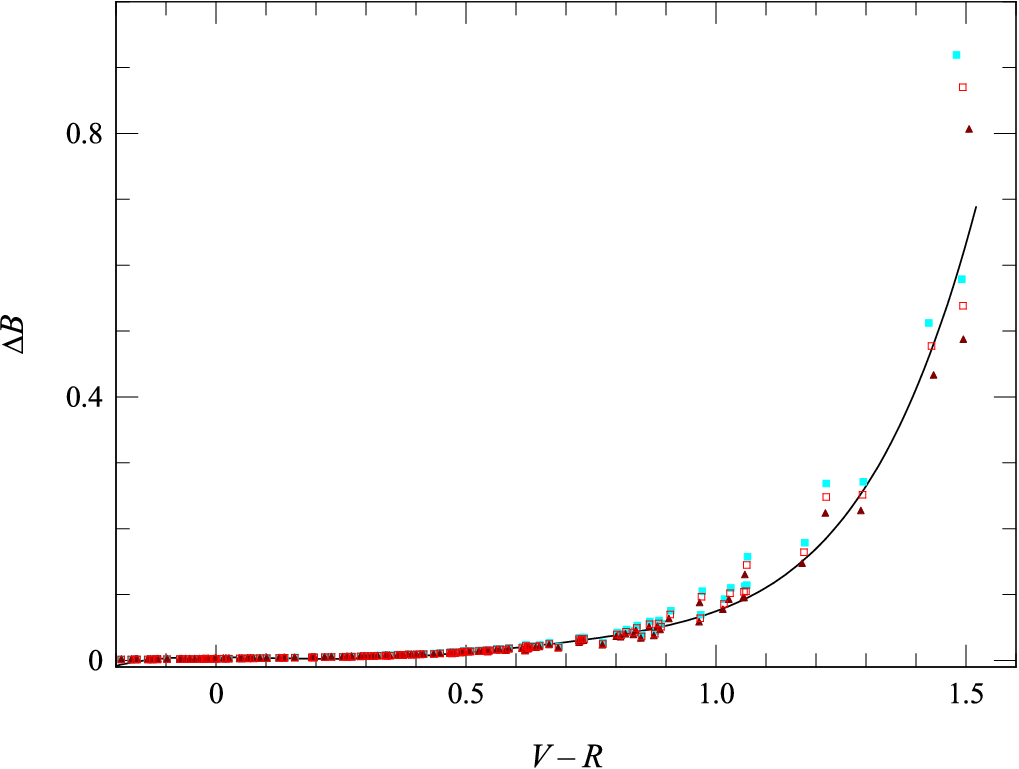}
}
\caption{Dependence of the red leak value in the $B$ band on the color indices for different values of $PWV$ at $Mz=1.15$. The notations are the same as in Fig.~\ref{fig:fig_dU_pwv}
}
\label{fig:fig_dB_pwv}
\end{figure}

There are at least two other factors affecting the observed energy distribution in the spectra of stars~--- the change in the absorption of radiation by the Earth's atmosphere due to changes in the object's altitude above the horizon and the change in the amount of water vapor. The effect of these factors on the red leak value is shown in Fig.~\ref{fig:fig_dU_alt} and Fig.~\ref{fig:fig_dU_pwv} for the $U$ band and in Fig.~\ref{fig:fig_dB_alt} and Fig.~\ref{fig:fig_dB_pwv} for the $B$ band. It can be seen that in the $U$ band the influence of PWV is practically not noticeable, while in the $B$ band this influence is present and can reach $\sim 0.1^m$ for red stars. 

\section{Photometric monitoring of symbiotic stars}
\mbox{}\vspace{-\baselineskip}

Photometric monitoring surveys to search for and study the rapid variability of symbiotic stars in the CMO are periodically carried out with the RC600 telescope. They usually take 2-4 hours (with a typical variability time of 10-60~min, see, e.g., \cite{Maslennikova2022}, \cite{Maslennikova2023}), during which CCD-frames are taken continuously (with an interval for the readout time, which for the \textit{Andor iKON-L} matrix is 4~s) of the sky area around the star under study. Brightness estimates are made by aperture photometry by comparing the signals accumulated in each frame from the variable star and from the comparison stars. It is considered that the comparison stars should have a spectral type close to that of the star under study~--- this avoids some problems in reducing the observations. In the case of observations of symbiotic stars, this requirement is difficult to fulfill because one of the components of a symbiotic star is usually a red giant, and such stars may not be in close proximity to the variable. Therefore, all stars in the field with a sufficient signal-to-noise ratio serve as comparison stars. 

The fast brightness variability of a symbiotic star or flickering~--- a chaotic change in the star's luminosity associated with an accretion disk or hot component whose emission contributes noticeably to the total flux from the system in the $U$ and $B$ bands and is almost invisible in the longer wavelength bands dominated by the emission from the red giant. The flux from the red giant can be considered constant on characteristic flicker effect timescales, so the additive signal from the red leakage remains constant in observations of fast light variability. However, many symbiotic stars have active and quiescent states in which the accretion rate is different, i.e., the contribution of the accretion disk to the system's light in the short-wavelength bands changes. Thus, in the blue region, the spectrum of a symbiotic star is very different from that of normal stars, so the plots presented above, which are intended to correct observations of normal stars, cannot be used directly to account for red leakage.

In general, to calculate the red leak value, it is necessary to know the energy distribution in the red region (spectral type or color index), the magnitude of $E(B-V)$, the brightness in the $R_c$ or $I_c$ band, and the altitude of the object at the time of the observation. Using these data and the bandwidth of the filter, camera entrance window, and detector, the number of red leak photons for a given object is calculated. The object's brightness measured during the observations in the $U$ or $B$ bands is converted to a photon flux from which the calculated red leak is subtracted. The flux is then converted back to stellar magnitudes.

For the case where the object is a normal star in the red part of the spectrum (e.g., a red giant in a symbiotic system), we can use the data in Fig.~\ref{fig:fig_dU}~-- \ref{fig:fig_dB_pwv} with a correction for the difference in the brightness of the system in the $U$ (or $B$) band from the spectral class expected for a normal star, coinciding with that of the red giant.

\begin{figure}[h]
\center{
\includegraphics[width=0.45\linewidth]{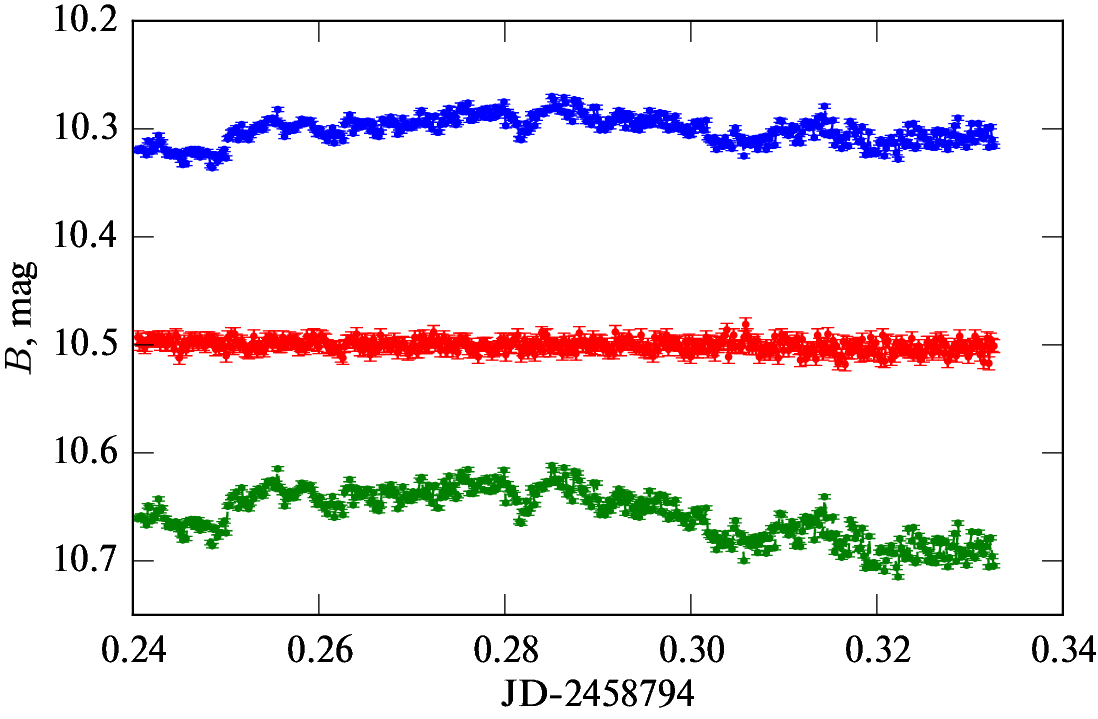}
\includegraphics[width=0.45\linewidth]{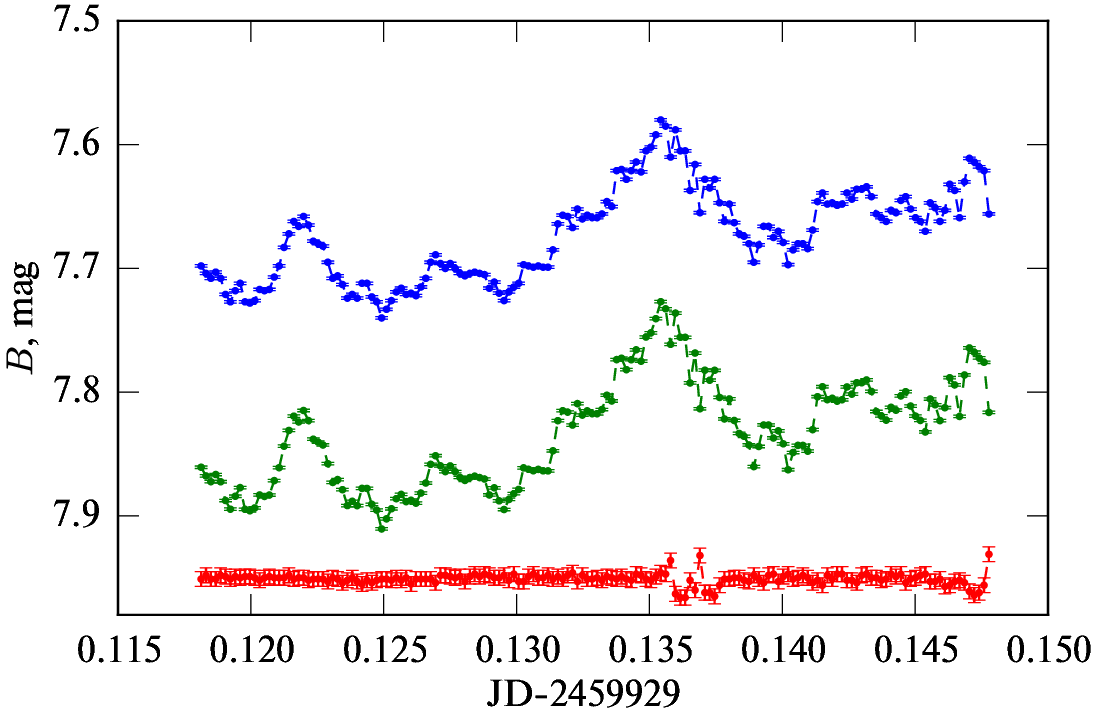}
}
\caption{Light curves of CH~Cyg in the $B$ band (observed~--- blue dots, corrected for red leak~--- green dots) and control star (red dots) obtained on 06.11.2019 (left) and 15.12.2022 (right). The control stars are $\sim 3^m$ fainter than the variable
}
\label{fig:fig_CH_Cyg}
\end{figure}

\begin{figure}[h]
\center{
\includegraphics[width=0.5\linewidth]{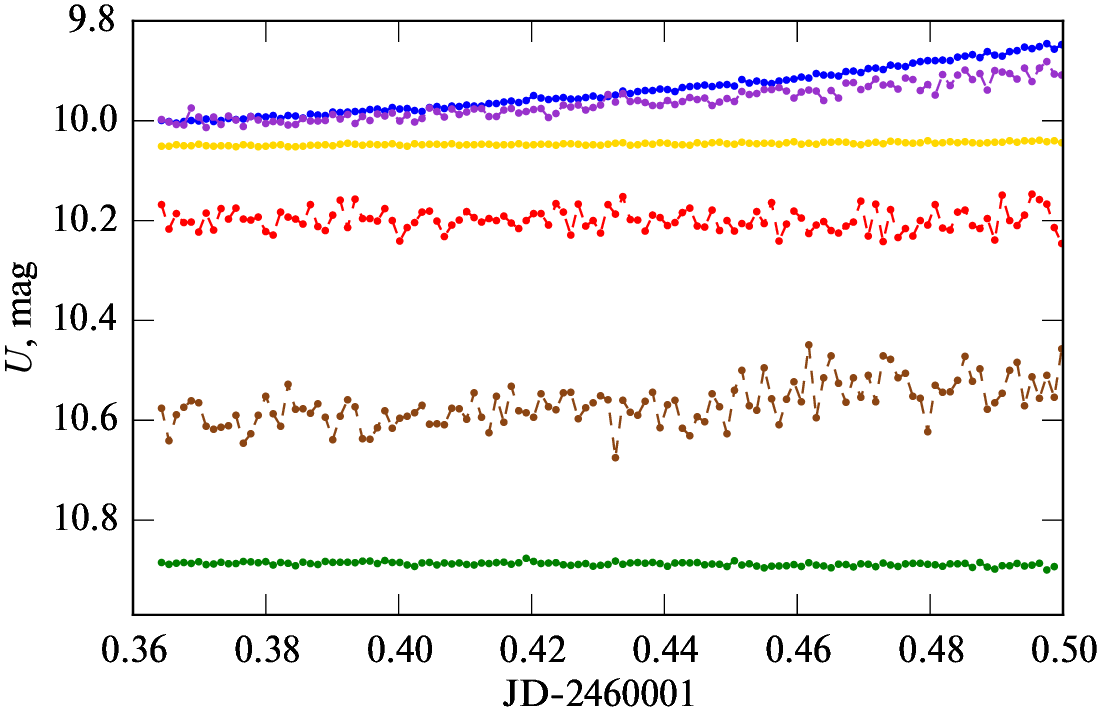}
}
\caption{Light curves of SU~Lyn in the $U$ band (observed curve~---blue dots, corrected for red leakage~---green dots) and comparison stars with color indices $V-R=0.75$ (purple dots), $V-R=0.63$~---brown dots, and $V-R<0.3$~--- red and yellow dots. The curves of the comparison stars are arbitrarily shifted along the vertical axis, with fainter stars showing a larger dispersion of the brightness estimates
}
\label{fig:fig_SU_Lyn}
\end{figure}

Fig.~\ref{fig:fig_CH_Cyg} shows the results of observations of the fast variability of CH~Cyg in the $B$ band made on November 6, 2019 and December 15, 2022. The duration of the first monitoring was 2h 12m, the second~--- about 45 m. On both light curves, irregular  brightness fluctuations are clearly visible. The maximum amplitude was $~0.07^m$ and $0.17^m$, respectively, with the RMS of the control star brightness being $0.006^m$ and $0.005^m$. It is difficult to distinguish the characteristic time of variability in the 2019 curve, but several short-term episodes of brightness change can be noted. On the curve of December 15, 2022, the same short-term light rises with an amplitude of $\sim 0.1^m$, separated by a time interval of $~20$~minutes, and low-amplitude flicker noise between them are clearly visible. 

To account for the red leak in the CH~Cyg light curves, the following system parameters were taken: spectral class of the cold component CH~Cyg~--- M7III \cite{Shenavrin2011}, $E(B-V)=0$ \cite{Green_2019}. The brightness of CH~Cyg in the $R$ band according to the \cite{Kloppenborg2023} was $\sim7.2^m$ on 11/06/2019, and $\sim7.2^m$ on 12/15/2022~--- $\sim5.3^m$. The object's altitude during the observations varied in the range 51-25$^\circ$. 

The corrected light curves are shown in Fig.~\ref{fig:fig_CH_Cyg}. Although the cool component of CH~Cyg was $2^m$ fainter in 2019 than in 2022, the average red leak on 11/06/2019 was higher~---it was about $0.4^m$, and on 12/15/2022~--- $0.17^m$. This is because the contribution of the hot component in the $B$ band in 2019 was much lower (by $\sim 3^m$), resulting in a large relative leakage contribution. After accounting for the leakage, the amplitude of the fast brightness variability increased on 06.11.2019 by 30\% (from $0.07^m$ to $0.10^m$), and on 15.12.2022 by about 10\% (from $0.17^m$ to $0.19^m$). The brightness of CH~Cyg in the $B$ band became fainter ($10.66^m$ and $7.83^m$, respectively). From Fig.~\ref{fig:fig_CH_Cyg} (left panel) we can see that the leakage correction allowed us to identify a dip in the light curve at $\Delta JD > 0.28$, blurred by the additional red leakage associated with the increase in $M_z$ during the monitoring.

The light curves of CH~Cyg are very different (Fig.~\ref{fig:fig_CH_Cyg}). On 12/15/2022, compared to 11/6/2019, the $B$-band brightness increased by $\sim 3^m$ and the amplitude of the flickering increased by a factor of 2. The $B-V$ color index decreased from $1.8^m$ to $1.3^m$ \cite{Kloppenborg2023}, indicating an increased contribution of the accretion disk to the system's brightness. Despite the fact that CH~Cyg is a double system with an orbital period of 15.6 years and a red giant pulsation period of 750 days \cite{Hinkle_2009}, we cannot explain these changes by the eclipse of the hot component, since on 06.11.2019 the orbital phase was 0.19, and on 15.12. 2022~--- 0.39. In the quiescent state, the spectrum of CH~Cyg is dominated by the emission of the cool component~--- the red giant; the contribution of the accretion disk and nebula in the blue region of the spectrum becomes noticeable only when the accretion rate increases \cite{Tarasova_2021}. Therefore, we can assume that the accretion disk had a small luminosity on 06.11.2019, and the observations on 15.12.2022 occurred during a period of significantly increased accretion rate, so the contribution of the accretion disk to the system's luminosity in the $B$ band increased and the amplitude of the flickering increased.

Fig.~\ref{fig:fig_SU_Lyn} shows the result of SU~Lyn brightness measurements for more than 3 hours. It is well seen that there are no oscillations on the light curve, but there is an almost linear trend~--- over 200 minutes the light has increased by $0.15^m$. A trend can also be seen in comparison stars with a large color index: for the star with $V-R=0.75$, the light changes were $0.10^m$, for the star with $V-R=0.63$~--- $0.09^m$. The brightness of the comparison stars with small $V-R$ color indices did not show any trend. During the observation period, the star's altitude changed from $57^\circ$ to $30^\circ$. 

The SU~Lyn spectrum matches well with the spectrum of the red giant M5III \cite{Mukai_2016} even in the blue region. Thus, the contribution of the accretion disk, if the disk is present in the system, can be neglected in the $U$ and $B$ bands. Therefore, the plots shown in Figs. ~\ref{fig:fig_dU} and \ref{fig:fig_dU_alt} were used to correct the red leakage. The magnitude of the red leak for SU~Lyn ranged from $0.88^m$ to $1.04^m$ depending on the height of the object. After taking into account the red leak, the light curve trend disappeared (see Fig.~\ref{fig:fig_SU_Lyn}), which may indicate the correctness of the correction.

\section{Conclusion}
\mbox{}\vspace{-\baselineskip}

The paper considers the manifestation of the "red leak" effect {} of the short-wavelength filters $U$ and $B$ in photometric observations with the CCD camera of the RC600 telescope of the CMO of SAI. The presence of transmission outside the main band of the filters leads to a significant contribution of this effect~--- up to $0.6-0.8^m$ for normal stars of late spectral types. For stars of early spectral types, the effect is small for the $B$ band, but can be noticeable in the $U$ band, where it is noticeable at the level of a few percent starting with A0-stars.

Fig.~\ref{fig:fig_dU} shows that the red leak in the $U$ band appears differently for stars having the same color indices but belonging to different luminosity classes. No such separation is observed in the $B$ band. This difference may be due to the shape of the filter transmission curves outside the working band~--- in the $U$ filter, part of the red leak signal is collected in a narrow transmission peak in the 700~nm region, whereas in the $B$ band, the entire spurious signal is collected in the IR region of the spectrum (see Fig.~\ref{fig:fig01}). When processing photometric observations and selecting comparison stars, we usually deal only with color indices, paying no attention to either the actual spectral type or the luminosity class of the comparison stars used in photometry. However, when working in the $U$ band, this can lead to significant errors when correcting for the effect of the red leak of the filter (the magnitude of the effect differs by a factor of 2 for stars of different luminosity classes).

The red leak contribution to the final brightness of the system is influenced by several effects that change the ratio of the fluxes in the short-wavelength bands relative to the long-wavelength bands. This can be interstellar reddening, which leads to the weakening of the object in the $U$ and $B$ bands and sharply increases the red leak contribution, which for normal stars at $E(B-V)=1$ can reach $2^m$, i.e., the spurious signal is several times larger than the useful one (see Figs. ~\ref{fig:fig_dU_EBV} and \ref{fig:fig_dB}).

The dependence of the magnitude of the "red leak"{} in the bands $U$ and $B$ on the interstellar reddening is explained by the strong fall of the radiation flux in the working transmission region of short-wavelength light filters relative to the flux in the bands $V$, $R$, and $I$ due to the action of absorption. For example, for the normal law of interstellar reddening at $E(B-V)=1$, the absorption will be $~2^m$ larger in the $U$ band than in the $R$ band, from which most of the leakage signal is collected. Therefore, the absolute magnitude of the leak will become smaller (since there is also absorption in the red region, although smaller in magnitude), with a significant increase in its relative contribution.

The value of the "red leak" {} is also affected by the absorption of radiation in the Earth's atmosphere. In this work, synthetic photometry was carried out for several characteristic values of the factors that produce this absorption~--- the change in PWV and the change in the airmass $M_z$. The first factor has little or no effect in the $U$ band, but is noticeable in the $B$ band. The second factor has a stronger effect in the $U$ band. The influence of the airmass is confirmed by observations of the red variable SU~Lyn (spectral type M5III)~--- the observed effect is in good agreement with the model effect (Fig.~\ref{fig:fig_SU_Lyn}).

In this paper, we propose methods to account for the red leak for stars whose spectrum is close to that of normal stars or differs in the blue region. In the first case, the accounting can be done directly using the data of Fig.~\ref{fig:fig_dU}~-- \ref{fig:fig_dB_pwv}. In the second case, the correction algorithm is more complicated and is described in the section "Photometric monitoring"{}.

In this paper, we present the results of the search for fast variability in two objects~--- the known pecular symbiotic star CH~Cyg and the recently discovered symbiotic star SU~Lyn. No flicker effect was detected in SU~Lyn. When the red leak is taken into account, the trend in its light curve disappeared, which confirms the correctness of the correction. The importance of accounting for the red leak is also confirmed by the observations of the flickering in CH~Cyg. The amplitude is an important parameter characterizing the rapid variability of the brightness. After the red leak correction, it increased to $0.10^m$ on 06.11.2019 and $0.19^m$ on 15.12.2022. After the correction, the star's brightness became lower: on 06.11.2019 in the $B$ band, the brightness decrease was $\sim 0.4^m$, and on 15.12.2022 the brightness dropped by $\sim 0.2^m$. The different magnitude of the correction (with an almost unchanged spectral type of the cool component) is due to the different contribution to the total brightness of the accretion disk system on these dates. It is also interesting to note that, despite the later spectral type of the cool component CH~Cyg compared to SU~Lyn, the correction for it is smaller. This is due to the presence in CH~Cyg of additional emission in the short-wavelength region, which is absent in SU~Lyn.

\bigskip

The work was performed using equipment purchased under the Lomonosov Moscow State University Development Program. The work of A. M. Tatarnikov (analysis of observational results and synthetic photometry) was supported by the Russian Science Foundation (grant 23-12-00092). N. A. Maslenikova (observations of symbiotic stars, processing of observations of rapid variability) is grateful for support from the Fund for the Development of Theoretical Physics and Mathematics <<BAZIS>> (grant 22-2-10-21-1).


\newpage

\bigskip\footnotesize

\end {document}